\documentstyle[aps,epsf,floats,twocolumn,prl]{revtex}
{\bf }

\begin{document}
\twocolumn[\hsize\textwidth\columnwidth\hsize\csname@twocolumnfalse%
\endcsname

\title{Permutation-Symmetric Multicritical Points in Random Antiferromagnetic
Spin Chains.}
\author{Kedar Damle$^1$ and David~A.~Huse$^2$}
\address{$^1$Physics Department, Harvard University, Cambridge, MA 02138\\
$^2$Physics Department, Princeton University, Princeton, NJ 08544
}
\date{July 9, 2002}
\maketitle

\begin{abstract}
{The low-energy properties
of a system at a critical point may have
additional symmetries not present in the microscopic Hamiltonian.
This letter presents the theory of a
class of multicritical points that provide an interesting example
of this in the phase diagrams
of random antiferromagnetic spin chains.
One case provides
an analytic theory of the quantum critical
point in the random spin-3/2 chain,
studied in recent work by Refael, Kehrein and Fisher
(cond-mat/0111295).}
\end{abstract}

\vskip 0.3 truein
]

Many of the interesting but poorly understood systems of
interest to current condensed-matter physics research are
quantum many-body systems with both
strong quenched randomness and strong interactions.
One class of such systems where there are some experimental results~\cite{Expt1}
and significant theoretical progress has been
possible~\cite{MaDaHu,DSF1,HyYaBhGi,HymYan,MonJolGol,DamMotHus} is
antiferromagnetic Heisenberg spin chains.
Much of the physics of these systems is captured
by the model Hamiltonian
\begin{equation}
{\cal H} = \sum_{i} J_{i}\vec{S}_i \cdot \vec{S}_{i+1} \; ,
\label{1HAF}
\end{equation}
where the operator $\vec{S}_i$ represents
a spin-$S$ at site $i$ of a linear chain.
The nearest-neighbor exchanges $J_i$ are all positive, may be
random, and may have an imposed ``dimerization'' $\delta$:
\begin{equation}
J_i=J[1+\delta (-1)^i]\exp{(R\eta_i)} \; ,
\label{ji}
\end{equation}
where $R$ measures the strength of the randomness and the
$\eta_i$ are random numbers drawn from a distribution with
mean zero, variance one, and all moments finite.

This simple-looking Hamiltonian encodes a variety
of low-energy behaviors depending upon the dimerization $\delta$, the
randomness $R$, and the value of the spin $S$.  For example, for
$S=1$, the undimerized chain ($\delta=0$) has
a quantum critical point at some intermediate value of $R$ that 
separates low-disorder Haldane and high-disorder 
Random Singlet (RS)
ground states~\cite{HymYan,MonJolGol}; this point is a multicritical
point in the $R$-$\delta$ plane at which three phases meet~\cite{Dam1}. 
In recent work, a related transition between low and high
disorder RS states was also seen numerically in undimerized $S=3/2$ chains~\cite{RefKehFis}.
Here we examine these
critical points in a larger context,
showing that they are but two members of a 
countably-infinite class of random multicritical points.  The low energy
statistical properties of these special points exhibit the symmetry of the
permutations of $N$ identical objects, ${\cal S}_N$,
although for $N>2$ this is {\it not} a symmetry of the system's bare
Hamiltonian.

\begin{figure}
\epsfxsize=\columnwidth
\centerline{\epsffile{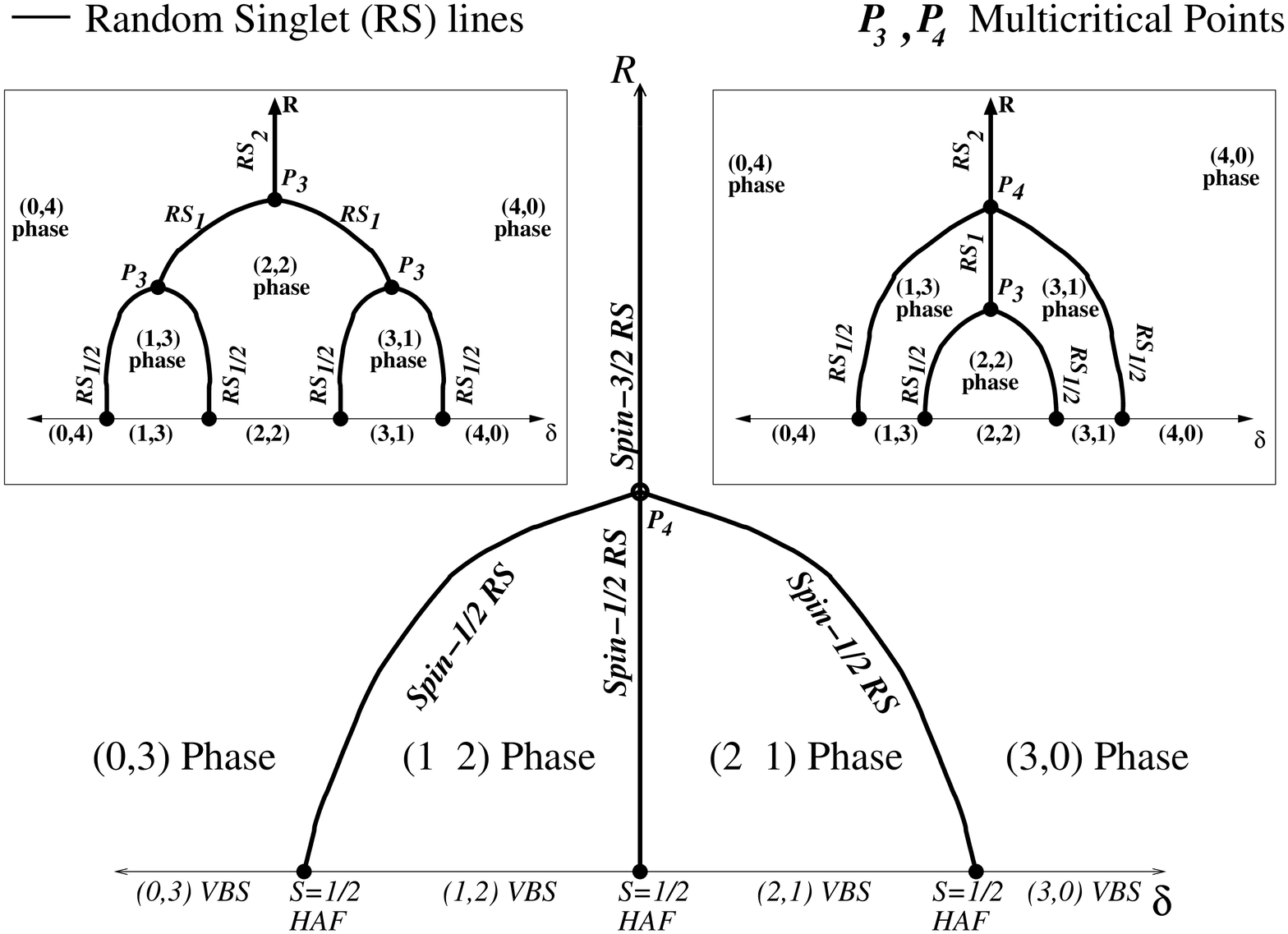}}
\vspace{0.15in}
\caption{Schematic phase diagram of $S=3/2$ chains in
the $R$-$\delta$ plane. Inset: Possible phases diagrams for $S=2$ chains.}
\label{phased3/2}
\end{figure}

To proceed, we describe the phases of the spin chain in the valence-bond
picture~\cite{AKLT} in which each
spin-$S$ is represented by the fully symmetrized multiplet of $2S$
spin-$1/2$'s.  The system's ground state has total spin zero
(modulo end effects), so each such spin-$1/2$ forms a 
singlet (a {\em valence bond}) with a
spin-$1/2$ on a neighboring site.  Thus, we can classify a ground state by
how many such valence bonds are formed across the even
links of the lattice: call this number $\sigma$.  Since each spin-$S$ must
participate in $2S$ valence bonds, there must be $(2S-\sigma)$
valence bonds across the odd links.  We will denote this {\em valence-bond
solid} (VBS) {\em ground-state}
as being in the ($\sigma$,$2S-\sigma$) {\em phase}, or, more compactly,
the $\sigma$ phase. 
For spin-$S$, there are (2S+1) such phases, and 
various phase transitions between them (some phase diagrams are shown in Fig~\ref{phased3/2}).
The VBS ansatz for the ground-state assumes that the valence bonds
are all between nearest neighbors, which is not precisely correct even
at $R=0$.
But there are indeed $2S+1$ topologically distinct possibilities
for the phases to which the real ground state can belong. As suggested
by the valence-bond description, these are
distinguished by the properties of a chain end:  For a chain in the
$\sigma$ phase, if an {\it even} bond in an infinite chain
is removed, the two resulting semi-infinite chains have ground
states that contain free spin-($\sigma/2$)'s localized near their ends.

The phase diagrams in the $R$-$\delta$ plane are simple for
zero or small $R$:
At $R=0$, all the phases
(with $\sigma = 0,1,\dots 2S$) can be accessed by sweeping
$\delta$ from -1 to +1, passing
a succession of $2S$ critical points~\cite{AffHal}; for integer $S$, the
($S$,$S$) phase that occurs around $\delta = 0$ is the familiar
Haldane phase~\cite{Hal1}, while for half-integer $S$, the critical point between $\sigma=S\pm (1/2)$
occurs at $\delta=0$.
At $R=0$, the low-energy properties of critical points separating phases
($\sigma$,$2S-\sigma$) and ($\sigma+1$,$2S-\sigma-1$)
arise from {\em residual spin-1/2's} obtained by first
forming $\sigma$ valence-bonds across the even links and $2S-\sigma-1$
valence-bonds across the odd links.  This leaves one unpaired spin-1/2
per site, and these spins behave as a (critical) spin-1/2 chain. 
Now, since the phases are gapped at $R=0$, they
survive for small $R$ as well. Likewise, the critical
points must extend to critical lines at $R \neq 0$, with the same
low-energy properties as the random-exchange
spin-1/2 chain~\cite{MaDaHu,DSF1}: Along these lines,
the chain is in the Spin-1/2
Random Singlet (RS$_{1/2}$) state. In this 
critical state, the {\em residual spin-1/2}
at any given site ``pairs'' into a singlet with one at some other site not
necessarily close to it, with the randomness
determining the pairing; this produces a glass of
single valence-bonds  with specific statistical properties at
low energies~\cite{DSF1}.

The behavior in the opposite limit of strong-randomness is
also readily understood. Due to the broad distribution of
exchanges at large $R$, the value of $S$ is not
crucial, and
a spin-$S$ version (RS$_S$) of the random singlet state obtains
for $\delta = 0$; at low energies, such a state is a glass of
$2S$-fold valence-bonds with the same statistical properties as the RS$_{1/2}$
state. When $\delta$ is non-zero and (say) positive,
the valence-bonds in the RS$_S$ state will have their
left end-point more often on an even site than on an odd site,
and the resulting state 
is in the ($2S$, $0$) phase.  Thus, turning on $\delta$ in this
regime drives the system into either the ($2S$, $0$) or  ($0$, $2S$)
phase, and the RS$_S$ state is the critical line
separating these two phases.

Given this picture of the phase diagram in the two limiting cases,
one is immediately led to the interesting possibility that
all the RS$_{1/2}$ critical lines meet the RS$_S$ line at a single point
at intermediate $R$ and $\delta=0$, producing a multicritical point at which
all $2S+1$ distinct phases of the spin-$S$ chain meet.
Indeed, a general theory of such multicritical points ${\cal P}_N$ at which
$N$ distinct phases meet forms the focus of
the present Letter.\narrowtext

We begin by addressing the question of {\em existence}:
Recent work~\cite{Dam1}
shows that the $\delta=0$ transition from
the gapless Haldane (1,1) phase to a RS$_1$
state in $S=1$ chains studied earlier~\cite{HymYan,MonJolGol}
is such a multicritical point (with $N=3$), where all three
phases of the system meet.  In the recent
RG study of the $S=3/2$ case with $\delta=0$, a single
quantum phase transition between a RS$_{1/2}$ state for small $R$
and a RS$_{3/2}$ state for large $R$ was observed
numerically~\cite{RefKehFis}. 
Our discussion above shows that this transition
is actually a multicritical point (with $N=4$), at which all four
distinct phases of a $S=3/2$ chain meet.
For $S=2$ or higher spin, the $N=2S+1$ multicritical point is
not generically present when one only varies $R$ and $\delta$,
to locate it requires tuning other parameters.  For $S=2$, in
particular, the possible topologies of the $R$-$\delta$ phase diagram
are shown in Fig~\ref{phased3/2}.

\begin{figure}
\epsfxsize=\columnwidth
\centerline{\epsffile{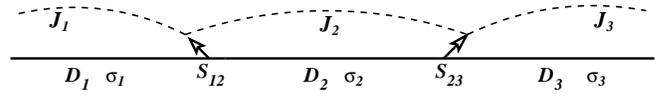}}
\vspace{0.15in}
\caption{A configuration of three adjacent domains.}
\label{dwallS}
\end{figure}

Such multicritical points represent points at which the local ground
state of the chain can be in any one of $N$ possible phases.
Thus, we develop a theory for the low-energy
physics of these points in terms of the domain structure
of the chain. We
begin with a low energy picture of the chain as being made up of
a sequence of domains, each belonging to one of the $N$
possible phases of a spin-$S$ chain ($N \equiv 2S+1$).
Neighbouring domains are separated
by {\em domain walls}. These domain walls each carry spin:
The magnitude of the spin
on a wall is given by the number of unpaired
spin-1/2's due to the difference of $\sigma$ across
the wall.  For two adjacent domains $D_1$ and $D_2$, with
$\sigma_1$ and $\sigma_2$, as in Fig~\ref{dwallS}, the
spin on the domain wall separating them has magnitude
$S_{12} = |\sigma_1 - \sigma_2|/2$.

The low-energy properties of our chain
are now controlled by the effective exchange couplings between neighboring
domain-wall spins. In the absence of neighboring domain-walls, each
domain-wall spin represents a zero-energy multiplet, with the spin
localized near the wall. Neighboring domain-wall spins
thus interact with an effective exchange $J$ that falls off 
rapidly with the domain length,
and is consequently broadly distributed in magnitude
(it can be of either sign).
We allow each type of domain, $\sigma$, to have its own probability distribution 
for the length of the domain and thus the exchange across the domain.
We thus have $N$ probability
distributions $P_{\sigma}(\beta|\Gamma)$ for
the corresponding log-couplings $\beta \equiv \ln(\Omega/|J|) \geq 0$,
where $\Omega$ is the cutoff energy (the strongest exchange), and
$\Gamma \equiv \ln(\Omega_0/\Omega)$ with $\Omega_0$ a bare cutoff.

The signs of the exchanges $J_i$ in the domain picture
are dictated by the domain sequence.
Consider the configuration in Fig~\ref{dwallS}, assuming that $J_2$ is the
strongest of the three exchanges shown.
At energy below $|J_2|$ but above $|J_1|$ and $|J_3|$, 
it should be possible to describe
the system by replacing $S_{12}$
and $S_{23}$ with a single effective spin $S_{13}$ whose value
is determined by the ground-state multiplet of the two-spin Hamiltonian
$J_2\vec{S}_{12} \cdot \vec{S}_{23}$. Consistency requires that
this must be the same as eliminating $D_2$ and having
a direct domain wall between $D_1$ and $D_3$ carrying
spin $S_{13} \equiv |\sigma_1 - \sigma_3|/2$. For this to be true, $J_2$
must be antiferromagnetic (positive) if $\sigma_1 - \sigma_2$ and
$\sigma_3 - \sigma_2$ are of the same sign, and ferromagnetic (negative)
otherwise.\narrowtext

To proceed further,
we need to specify the probabilities with which different domain sequences
occur.
We do this within a nearest-neighbor transfer
matrix formalism. Thus we have a symmetric, purely off-diagonal $N \times N$
transfer matrix $W_{\sigma\sigma^\prime}$ which gives the relative
weights for the two types of domains $\sigma\neq\sigma^\prime$ 
to be present and adjacent to each other.
We normalize $W$ to make its largest eigenvalue $+1$
(we denote the components of the corresponding normalized eigenvector as
$\sqrt{\rho_\sigma}$). This guarantees that the `partition function'
(sum over all possible
domain configurations)
$Z_L \equiv {\rm Tr}(W^{L})$ for a sequence of $L$ domains with periodic
boundary conditions tends to unity as $L \rightarrow \infty$.
The unconditional probability for a given segment, say
$\dots \mu_a\mu_b \mu_c \dots$, to occur in the spatial sequence of domain-types
is now given by modifying the expression for $Z$ by 
introducing appropriate projection operators
$\Pi^{\mu}_{\sigma \sigma^{\prime}} \equiv\delta_{\sigma \sigma^\prime}
\delta_{\sigma \mu}$ at the corresponding places in the product of $W$'s,
yielding the expression
${\rm Tr}(\dots W\Pi^{\mu_a}W\Pi^{\mu_b}W\Pi^{\mu_c}W \dots)$. Thus,
the probability of the $k^{\rm th}$ domain being type $\mu$ is $\rho_\mu$,
that of the $k^{\rm th}$ domain being type $\mu$ {\em and} the 
$(k+1)^{\rm th}$ being type $\nu$ is $\sqrt{\rho_\mu \rho_\nu}W_{\mu \nu}$,
etc. (for $L \rightarrow \infty$).

Given the broad probability
distributions $P_{\sigma}$ of the log-couplings $\beta$ in our domain model,
we can analyze the low energy properties
using a {\em strong-disorder RG} approach that eliminates, at each step,
{\em all} excited states of the strongest-coupled pair
of remaining domain-wall spins~\cite{HymYan}.  
For our domain-wall model, the RG
action is rather simple, and
 in this strongly random limit
does not generate any correlations between domains beyond those
given by the nearest-neighbor transfer matrix $W$:
Consider Fig~\ref{dwallS} and let $J_2$ be the
strongest bond.  At each step,
this RG ``integrates out''
the domain (in this case $D_2$) straddled by the strongest coupling:
If $\sigma_1 \neq \sigma_3$, $D_2$ is eliminated,
a direct domain-wall between $D_1$ and $D_3$ is formed, and the
signs, but {\it not} the magnitudes, of $J_1$ and $J_3$ 
are altered (if necessary) to conform to
the requirements of the sign-rule for this new configuration. If
$\sigma_1 = \sigma_3$, $D_2$ is eliminated and $D_1$ and $D_3$
are merged together into one domain. 
This merged domain is straddled by a renormalized
coupling of magnitude $|J_{13}| = |J_1J_3/J_2|$ and sign determined by
our sign-rule.
A little book-keeping yields the RG flow equations
corresponding to this procedure:
\begin{eqnarray}
\frac{dW_{\sigma \sigma^\prime}}{d\Gamma}
& = & V_{\sigma \sigma^\prime} - \frac{W_{\sigma \sigma^\prime}}{2}
[P_\sigma^0 + P_{\sigma^\prime}^0 -
V_{\sigma \sigma}
-V_{\sigma^\prime \sigma^\prime}] \, ,
\nonumber \\
&&\!\!\!\!\!\!\!\!\!\!\!\!\!\!\!\!\!\!\!\!\!\!\!\!\!\!
\frac{\partial P_\sigma}{\partial \Gamma}
= \frac{\partial P}{\partial \beta} + P_\sigma^0P_\sigma(\beta|\Gamma)+
V_{\sigma \sigma}
(P_\sigma \otimes P_\sigma -P_\sigma) \, , \nonumber \\
\frac{dL}{d\Gamma} & = & -L[\rho \cdot P^0 + Y] \, , 
\end{eqnarray}
where $V_{\alpha \beta} \equiv 
\sum_{\mu}W_{\alpha \mu}P^0_{\mu}W_{\mu \beta}$,
$P_\mu^0 \equiv P_\mu(0|\Gamma)$, $P_\sigma \otimes P_\sigma \equiv \int \int 
d\beta_1 d\beta_2 P_\sigma(\beta_1|\Gamma)
P_\sigma(\beta_2|\Gamma)\delta(\beta-\beta_1-\beta_2)$,
$Y=\sum_{\nu}
\rho_\nu V_{\nu \nu}$, $\rho \cdot P^0 = \sum_{\mu}
\rho_\mu P^0_\mu$ and the sums run over the labels of the $N$ domain types.
Moreover, the flow of $W$ also induces a change in $\rho$:
\begin{eqnarray}
\frac{d\rho_\sigma}{d\Gamma} &=& -\rho_\sigma(P_\sigma^0 +V_{\sigma \sigma}
-\rho \cdot P^0- Y )\, .
\end{eqnarray}

We turn now to a fixed point analysis of these RG equations. The multicritical
point ${\cal P}_N$ is controlled by a fixed point with
${\cal S}_N$ statistical
symmetry corresponding to freely interchanging between the $N$ phases
that meet at this point. This fixed point has
$W_{\mu \nu} = 1/(N-1) \;\forall \; \mu \neq \nu$,
$P_\mu(\beta|\Gamma) = (N-1)e^{-(N-1)\beta/\Gamma}/\Gamma
\;\; \forall \; \mu$, and $\rho_\mu = 1/N \; \forall \; \mu$.
Also, the number of domains decreases with the cutoff as
$L(\Gamma) =  L(0)/\Gamma^{1/\psi_{N}}$, with the exponent
$\psi_N \equiv 1/N$. Thus, all $N$ domain types are equally likely
at this fixed point, with any two types
of domains equally likely to be adjacent to each other. 
The fractions $p_s$ of the domain walls in the low-energy effective
Hamiltonian that have spin-s follow simply from
this. In the $S=3/2$, $N=4$ case, we 
predict $1/\psi_4 = 4$, $p_{1/2} =1/2$, $p_1 = 1/3$ and
$p_{3/2} = 1/6$; the
numerical estimates of \cite{RefKehFis} are in reasonable agreement
with these results. [The low-temperature specific heat and
susceptibility at the critical point
are completely determined~\cite{DSF1,HymYan,MonJolGol,RefKehFis} by
$\psi$ and $p_s$.]

To get information on off-critical scaling properties, we need
to analyze small perturbations about this ${\cal P}_N$ fixed point.
Fortunately, the ${\cal S}_N$ symmetry imposes enough
structure on the linearized flows to allow a full calculation of all
RG eigenvalues $\lambda$ which govern the
$\Gamma^\lambda$ growth or decay of the corresponding eigen-perturbations:
There are only $N-1$ relevant eigenvectors, all having
eigenvalue $\lambda^+_{N} \equiv (\sqrt{4N+1} - 1)/2$.
Since one has to tune $N-1$ `knobs' in general to get
$N$ phases to all be `degenerate', this coincides with the minimal
possible number of relevant directions at a $N$-fold multicritical point---thus,
this ${\cal S}_N$ fixed point governs
{\em all} such strongly-random $N$-fold multicritical points. [In
contrast, usual (non-random) multicritical points
in Landau theory or in two-dimensions do not generically
have ${\cal S}_N$ low-energy symmetry. For example, at the critical
fixed point of the two-dimensional three-state Potts model, there are
four relevant modes, two of which produce flows to generic tricritical
points that do not have the ${\cal S}_3$ symmetry of the Potts model.]
The relevant
eigenvectors can be chosen to correspond to perturbations which
make only one of the $N$ phases fall out of favour, thus reducing the
symmetry from ${\cal S}_N$ to ${\cal S}_{N-1}$. 
In addition we have one irrelevant eigenvector with eigenvalue $-1$
(representing an additive shift in $\Gamma$), for $N>2$ there are
$N-1$ irrelevant eigenvectors with
eigenvalue $-(\sqrt{4N+1} +1)/2$, and for $N>3$ there are
$N(N-3)/2$ irrelevant eigenvectors with eigenvalue $-N$.
[We also expect~\cite{DSF1} other `infinitely' irrelevant perturbations,
{\em i.e.} decaying exponentially with $\Gamma$---these are not considered
here.] 

In the $S=3/2$, $N=4$ case, we thus predict
a relevant eigenvalue of $\lambda^+_4 = (\sqrt{17}-1)/2$, and
a correlation length exponent $\nu \equiv 1/\lambda^+_4\psi_4 \cong
2.56$ (note that the numerical estimate of
\cite{RefKehFis} differs significantly from this prediction, probably
due to slow transients or finite-size effects). Deviations from
${\cal P}_4$ in the $R$-$\delta$ plane contain linear combinations
of the three relevant perturbations.
The fact that the RG eigenvalues are
all the same means that the phase boundaries come in linearly
at ${\cal P}_4$. The slope of two
of the four phase boundaries is fixed by noting that
any phase boundary between phases related by the $\delta \rightarrow -\delta$
symmetry of the problem (corresponding to an interchange of
even and odd sites) must lie on the $R$ axis. The same symmetry forces the
the remaining two to be reflections of one another about the $R$ axis.
To fix their slopes, consider one of them,
say the (2,1) to (3,0) phase
boundary:  These two phases are degenerate here, but the
other two have higher energy.  A positive $\delta$ lowers
the energy of both phases (since they have more singlets
on the even bonds than on the odd bonds),
with the energy of the (3,0) phase lowered more
than that of the (2,1) phase.  Decreasing $R$, on the other hand, 
lowers the energy of the (2,1) phase relative to the (3,0) phase.
Thus, these two phases will remain degenerate if $R$
is {\it decreased} with increasing $\delta$, and this
phase boundary must leave ${\cal P}_4$ with a negative slope
(see Fig.~\ref{phased3/2}).

The above RG equations also admit fixed points of lower symmetry ${\cal S}_M$
(with $M < N$) at which domains of $M$ phases
each occur with equal probability,
and other domain-types do not occur at low energy. 
These fixed points govern locii of
multicritical points ${\cal P}_M$ at which $M$ of the phases meet.
Two examples
of such locii are the RS$_{1/2}$ and RS$_{3/2}$ lines in the 
$R$-$\delta$ phase diagram
of $S=3/2$ chains (these have $M=2$). Other examples (with $M=3$)
include the points ${\cal P}_3$ in the $R$-$\delta$ phase diagram
(see Fig~\ref{phased3/2}) of
$S=2$ chains---these have the same exponents as the point ${\cal P}_3$
of $S=1$ chains.
In addition to these lower-order multicritical points, 
the RG equations also admit `Griffiths' fixed points, 
which describe the
continuously varying power-law singularities within the individual
phases~\cite{Dam1}.

Turning to $S>3/2$, it is now clear that the generic phase
diagram in the $R$-$\delta$ plane for $S=2$ will look like one of the
two insets shown in Fig~\ref{phased3/2},
with the putative ${\cal S}_5$ symmetric
point ${\cal P}_5$ splitting into lower-order multicritical points
as shown. All five phases of the $S=2$ chain
will only meet at ${\cal P}_5$ upon fine-tuning some additional parameter
in the model (such as nearest-neighbor interactions more general than
simply the exchange $\vec{S}_i \cdot \vec{S}_{i+1}$). 
Similar considerations also rule out
the generic occurrence of such maximally
symmetric multicritical points in the $R$-$\delta$ plane for all $S>2$.

Finally, we note that the basic structure of the domain-wall model used here
can also be motivated
from a more microscopic argument with (\ref{1HAF}) as the
starting point.  Consider treating
(\ref{1HAF}) for arbitrary $S$ with a
generalization of the
approximate {\em extended Ma-Dasgupta-Hu} RG
approach~\cite{MonJolGol,RefKehFis}.  
[For general $S$, the procedure eliminates {\em all excited
states} of the most strongly coupled pair of spins if this coupling
is ferromagnetic, while taking care to eliminate
only the highest excited state if it is antiferromagnetic; the RG
rules for signs and magnitudes of couplings are as in
Ref~\cite{MonJolGol,RefKehFis}.]
Assign the formal domain
label $\sigma = 0$ ($\sigma = 2S$) to every even (odd)
bond of the unrenormalized Hamiltonian (\ref{1HAF}).
Eqn~(\ref{1HAF}) with this labeling is consistent with the rules for
domain-wall spins and signs of couplings in our domain-wall model; we
can therefore formally think of each $J$ of (\ref{1HAF}) as straddling a domain of the corresponding type.
Now, if the
renormalized Hamiltonian and choice of labels at a given stage of the
RG is consistent with this domain interpretation, it
is possible to relabel the couplings after each RG step to preserve
this property:
Let $J_2S_{12} \cdot S_{23}$ in Fig~\ref{dwallS} be the term with
the largest gap between lowest and highest energy states.
If $J_2 > 0$,
and neither $S_{12}$ nor $S_{23}$ are spin-1/2,
change $\sigma_2$
to $\sigma_2 + (\sigma_1-\sigma_2)/|\sigma_1-\sigma_2|$ after the next
RG step (which reduces both $S_{12}$ and $S_{23}$ by $1/2$).
If $J_2 > 0$ and
$S_{12}=S_{23}=1/2$,
attach the common label of $J_1$ and $J_3$ to the new coupling
$\tilde{J}_{13}$ that reaches across them after the next step (which
puts them into a singlet state).
In all other cases, there is no need to relabel any of the couplings
that remain after the next step. Although this RG procedure
is approximate, it is expected to be qualitatively accurate for low-energy
properties~\cite{HymYan,MonJolGol,RefKehFis}, and
the formal device above thus provides an
alternative route to our domain model.

We acknowledge useful discussions with I.~Affleck, D.~S.~Fisher and G.~Refael, and
the support of NSF-DMR grants
9981283, 9714725, \& 9976621 (KD), and 9802468 (DAH).

\end{document}